\documentclass[aps,prl,twocolumn,superscriptaddress]{revtex4-2}
\usepackage{graphicx,epstopdf,siunitx,braket,xspace,amssymb,dsfont,bm,amsmath,float,xcolor}
\usepackage{appendix}
\usepackage{braket}
\usepackage{xcolor}

\newcommand{\SiOx}[0]{SiO$_2$\xspace}
\newcommand{\Vbias}[0]{$V_\mathrm{B}$\xspace}
\newcommand{\Vturnon}[0]{$V_\mathrm{T}$\xspace}
\newcommand{\Vg}[0]{$V_\mathrm{G}$\xspace}
\newcommand{\Vsd}[0]{$V_\mathrm{SD}$\xspace}
\newcommand{\figRef}[2]{Fig.~\ref{fig#1}(#2)}
\newcommand{\figReflong}[2]{Figure~\ref{fig#1}(#2)}
\newcommand{\red}[1]{\textcolor{black}{#1}}
\begin{document}

\title{Control of threshold voltages in Si/SiGe quantum devices via optical illumination}

% Force line breaks with \\
\author{M. A. Wolfe}
\altaffiliation[These authors ]{contributed equally}
\affiliation{Department of Physics, University of Wisconsin-Madison, Madison, WI, 53706, US}
\author{Brighton X. Coe}
\altaffiliation[These authors ]{contributed equally}
\affiliation{Department of Physics, University of Wisconsin-Madison, Madison, WI, 53706, US}
\author{Justin S. Edwards}
\affiliation{Department of Physics, University of Wisconsin-Madison, Madison, WI, 53706, US}
\author{Tyler J. Kovach}
\affiliation{Department of Physics, University of Wisconsin-Madison, Madison, WI, 53706, US}
\author{Thomas McJunkin}
\altaffiliation[Present address: ]{Johns Hopkins University Applied Physics Laboratory, Laurel, Maryland 20723, USA}
\affiliation{Department of Physics, University of Wisconsin-Madison, Madison, WI, 53706, US}
\author{Benjamin Harpt}
\affiliation{Department of Physics, University of Wisconsin-Madison, Madison, WI, 53706, US}
\author{D. E. Savage}
\affiliation{Department of Material Science and Engineering, University of Wisconsin-Madison, Madison, WI, 53706, US}
\author{M. G. Lagally}
\affiliation{Department of Material Science and Engineering, University of Wisconsin-Madison, Madison, WI, 53706, US}
\author{R. McDermott}
\affiliation{Department of Physics, University of Wisconsin-Madison, Madison, WI, 53706, US}
\author{Mark Friesen}
\affiliation{Department of Physics, University of Wisconsin-Madison, Madison, WI, 53706, US}
\author{Shimon Kolkowitz}
\altaffiliation[Present address: ]{Department of Physics, University of California-Berkeley, Berkeley, California 94720, USA}
\affiliation{Department of Physics, University of Wisconsin-Madison, Madison, WI, 53706, US}
\author{M. A. Eriksson}
\affiliation{Department of Physics, University of Wisconsin-Madison, Madison, WI, 53706, US}
\email{maeriksson@wisc.edu}

\begin{abstract}

% \textcolor{red}{Optical } illumination is widely used to “reset” quantum-dot qubit devices at cryogenic temperatures. While highly effective, this technique has not been well studied and is referred to as poorly understood lore.
Optical illumination of quantum-dot qubit devices at cryogenic temperatures, while not well studied, is often used to recover operating conditions after undesired shocking events or charge injection. Here, we demonstrate systematic threshold voltage shifts in a dopant-free, Si/SiGe field effect transistor using a near infrared (780 nm) laser diode. We find that illumination under an applied gate voltage can be used to set a specific, stable, and reproducible threshold voltage that,  over a wide range in gate bias, is equal to that gate bias. Outside this range, the threshold voltage can still be tuned, although the resulting threshold voltage is no longer equal to the applied gate bias during illumination. We present a simple and intuitive model that provides a mechanism for the tunability in gate bias. The model presented also explains why cryogenic illumination is successful at resetting quantum dot qubit devices after undesired charging events.

% \vspace{5em}
\end{abstract}

\maketitle

\section{I. Introduction}

Gate-defined semiconductor quantum-dots devices provide a platform for a wide variety of solid-state qubits~\cite{Hanson2007, Zwanenburg2013, Zhang2018, Burkard2023}. These devices are based on voltage-biased gates fabricated on top of semiconductor heterostructures, either epitaxial or oxide-on-semiconductor, and they provide a highly controllable electrostatic environment for electron and hole-based qubits. However, unlike their classical counterparts, which are substantially uniform across a chip, state-of-the-art quantum-dot devices show variation in operating points between gates even very close together on a chip~\cite{Neyens2023,meyer2023_3rd}.  Operating points can be manipulated by charge injection~\cite{Meyer2023, Massai2023} and bias cooling~\cite{Ferrero2024}. One potential source of variability between devices is at the gate-oxide interface, which hosts charge traps~\cite{Goetzberger2003} that lead to device instability, electrostatic disorder~\cite{Mi2015, Esposti2022}, and charge noise~\cite{Connors2019,Struck2020}. The latter is a significant source of dephasing for quantum-dot qubit operations~\cite{Dial2013, Kawakami2016, Thorgrimsson2017}, to the extent that sweet spots~\cite{Bertrand2015, Russ2015}, symmetric operating points~\cite{Reed2016reduced, Shim2016, Martins2016, Noire2022, Xue2022}, and other favorable points in the energy dispersion~\cite{Kim:2014p70} are used to mitigate its effects.  

\begin{figure*}
\includegraphics[width = \textwidth]{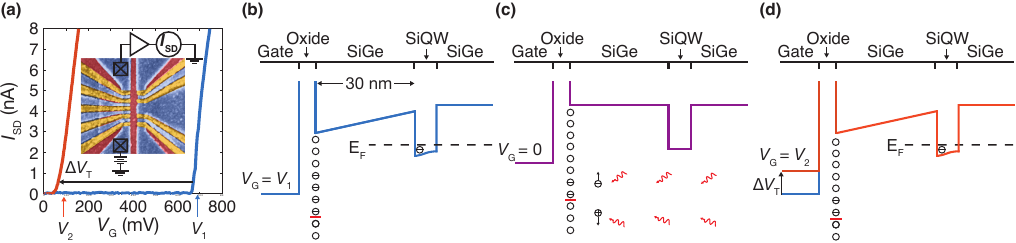}
\caption{\label{fig1} (a) Global turn-on curve of a quadruple quantum-dot device before (blue) and after (red) illumination. (Inset) A false-color scanning electron microscope (SEM) image of a Si/SiGe quadruple quantum-dot gate-defined device. Different colors denote the three different gate layers of the device, where all gates are treated as one global accumulation gate at voltage $V_\mathrm{G}$. Relevant ohmic contacts and measurement circuit are labeled. (b,c,d) show schematic band diagrams of the device, related to the turn-on curves in (a): (b) before reset (blue), (c) during reset (purple), and (d) after reset (red). $V_1$ and $V_2$ are voltages slightly above the threshold $V_\mathrm{T}$ needed to accumulate the same charge density in the quantum well, and $V_2 = V_1 + \Delta V_\mathrm{T}$}
\end{figure*}

\red{Threshold voltages for quantum-dot devices are difficult to predict upon cool down, with variations as large as 700~mV possible for nominally identically devices; in part, this variation can arise from trapped charge at the interface. A common experimental technique to reduce this variation is to illuminate the device with light, with a single-photon energy that is larger than the bandgap of the host semiconductor}~\cite{Wilamowski2001,Tyryshkin2005,Thalakulam2010,Thalakulam2011,Payette2012,Ward2013,Kim2017,Shetty2022}. The incident photons generate electron-hole pairs that, evidently, allow for the rearrangement of unwanted charge, which is otherwise locked in place at low temperatures. However, the processes, mechanisms, and limitations of this technique are not well understood.

Here we present a method to systematically tune the threshold voltage of dopantless Si/SiGe devices using optical illumination in the presence of an applied gate voltage. We show that such biased illumination provides precise \emph{in-situ} tuning of threshold voltages over a wide voltage range. We present a model that explains these results in terms of control of the density of trapped charge at the oxide-semiconductor interface. At large positive bias voltages, we argue that this method fills all the available interface states.  Under even larger applied gate biases, above \SI{1.5}{V} in the device studied here, Fowler-Nordheim tunneling results in metastable trapping of even more charge. Unlike measurements at positive gate bias, measurements at large negative bias voltages depend quadratically on light intensity, which suggests a two-photon process may be important in that regime. In addition to providing a tool for tuning threshold voltages, these results enable an understanding of illumination at cryogenic temperatures, which is widely used to provide a consistent reset for Si/SiGe quantum devices.

\section{II. Measurements}

\subsection{A. Overview}

To introduce the concept, we first demonstrate the ability to reset a Si/SiGe quantum-dot device at cryogenic temperatures using \emph{in-situ} illumination. The inset in \figRef{1}{a} shows a scanning electron micrograph (SEM) image of a Si/SiGe quadruple quantum-dot device with two charge sensors, lithographically identical to the device measured in this work. A global turn-on curve, where all gate voltages are swept simultaneously at the same voltage \Vg, is measured immediately after cooling down the device to \SI{1.2}{K} with a source-drain bias (\Vsd) of \SI{50}{\micro V}, as shown in \figRef{1}{a}. We define the threshold voltage (\Vturnon) to be the \Vg that achieves a source-drain current $I_{\mathrm{SD}}=\SI{1}{nA}$. The blue curve in \figRef{1}{a} corresponds to the initial turn-on curve with \Vturnon = \SI{675}{mV}. \red{After cryogenic illumination with light of wavelength \SI{780}{nm} from a laser diode (U.S. Lasers Inc D7805I) biased with \SI{15}{mA} of current, \Vturnon is dramatically reduced by \SI{600}{mV} to \SI{75}{mV} (red curve).}

Figures~\ref{fig1}(b)-(d) present schematic band diagrams of a gated Si/SiGe quantum well device before, during, and after illumination, illustrating how optical illumination can shift the threshold voltage of such devices. \red{In \figRef{1}{b}, trapped charge (filled circles) initially resides at the semiconductor-oxide interface, which in this case is formed of SiGe/Al$_2$O$_3$ (see Appendix A for details of the device fabrication)}. Here, the short red line indicates mid gap, and filled states above this level trap negative charge at the interface, whereas empty states below midgap trap positive charge (holes) at the interface~\cite{Sze2006}. This negative charge influences the voltage required to turn on the device, and as shown in \figRef{1}{a}, the threshold voltage in this case is $V_{\mathrm{T}} = \SI{675}{mV}$.
% Based on data and arguments discussed below in Sec.~B, we show in \figRef{1}{c}, corresponding to the system during illumination, a situation where there is a sufficient density of electron-hole pairs that the electric field is completely screened by charge at the oxide-semiconductor interface under the applied bias \Vg = \SI{0}{mV}.  In this case, the required charge density is negative and of smaller magnitude than in \figRef{1}{b}, and that new charge density is trapped at the interface when the illumination is terminated. 
During illumination at \Vg = \SI{0}{mV}, as shown in \figRef{1}{c}, photo-generated electron-hole pairs enable charge to accumulate as needed to make the electric field zero in the semiconductor heterostructure, as indicated by the flat bands. In the case shown, the charge density required to screen the electric field from the semiconductor region is negative and of smaller magnitude than the original charge density shown in \figRef{1}{b}. Following the illumination, a reduced charge density is trapped at the interface, as shown in \figRef{1}{d}, leading to a lower threshold voltage $V_{\mathrm{T}}=\SI{75}{mV}$ for accumulation. Repeating this reset procedure, i.e., illuminating again with \Vg = \SI{0}{mV}, causes no further shifts in \Vturnon \red{(see Appendix C for data on repeatability and stability).}

In the following we analyze a series of experiments in which a heterostructure field-effect transistor (H-FET), nominally identical to that shown in \figRef{2}{a}, is illuminated under a wide range of gate bias conditions at a temperature of $\SI{3}{K}$. \red{The heterostructure stack for the device is shown in \figRef{2}{b}. Here, the gate oxide is formed by dry oxidation, and additional details on device fabrication are given in Appendix A.} In the following, we observe markedly different results after illumination depending on the sign and magnitude of the bias voltage applied during illumination, which we present below in Secs.~B, C, and D.

% The final result is shown in \figRef{1}{d}, where the density of trapped negative interface charge is smaller than was initially present, allowing the device to accumulate a 2DEG at a lower threshold voltage ($V_{\mathrm{T2}}=\SI{75}{mV}$). Repeating this procedure, i.e., illuminating again with \Vg = \SI{0}{mV}, causes no further shifts in \Vturnon.

\subsection{B. Biased illumination at small gate bias}

\begin{figure}
\includegraphics[width = \columnwidth]{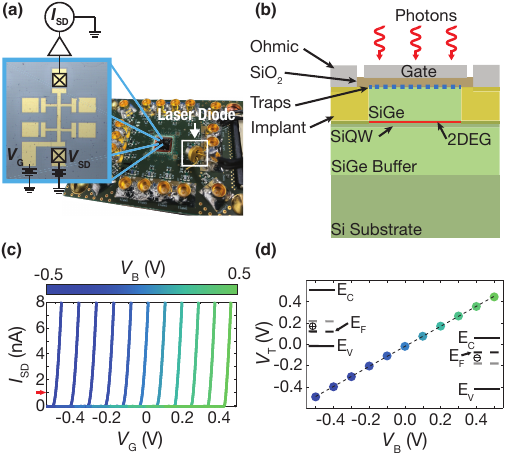}
\caption{\label{fig2} 
(a) Image of a typical circuit board and laser diode used for illumination and measurements of a Si/SiGe quantum device. (Inset) Optical micrograph of a H-FET used in the experiment with relevant ohmic contacts connected to the measurement circuit labeled. (b) Schematic illustration of the Si/SiGe H-FET device stack indicating locations of the two-dimensional electron gas (2DEG), trapped interface charge, and incident light radiation. (c) Measured device source-drain current after subsequent gate-biased illuminations for a range of bias voltages. The threshold voltage \Vturnon of each curve is determined at the current value $\mathrm{I_{SD}}$ = \SI{1}{nA} (red arrow). (d) Extracted \Vturnon as a function of gate voltage. The uncertainty in \Vturnon is smaller than the data points shown. Black dashed line is a linear fit, with a slope of 0.94. (Inset) Diagrams indicating how the position of the Fermi level relative to midgap (dashed gray line) determines the sign of the trapped charge.  
}
\end{figure}

Figure~\ref{fig2}(c) reports the turn-on curves following a series of illuminations at a gate bias of \Vbias. To ensure a consistent starting condition, before performing each gate-biased illumination we initialize (``reset'') the device by illuminating at a \Vbias of \SI{0}{V} using a laser diode current of \SI{15}{mA} for a duration of \SI{30}{s}. For each curve in Fig.~\ref{fig2}(c), we then illuminate  with the same laser current and duration and with a non-zero \Vbias, as indicated by the color bar. We then measure the turn-on curve using a \SI{1}{mV} source-drain bias. (We note that the data shown Fig.~\ref{fig2} appear unchanged even without the reset step, provided sufficiently long biased-illumination is performed.)

\figReflong{2}{c} shows that the turn-on curves shift dramatically as a function of the applied bias during illumination. The threshold voltage \Vturnon extracted from this data depends linearly on \Vbias with \red{a slope of $0.94\pm 0.01$}, a value very close to unity (\figRef{2}{d}).  We argue here that this behavior arises from mobile, photo-generated electron-hole pairs that, during illumination, move to screen the electric field in the semiconductor.  That is, after sufficiently long illumination, carriers of the correct sign accumulate at the oxide-SiGe interface in order to screen the electric field arising from both the applied bias voltage and the work function difference between the top gate and the electrical connection at the quantum well. Evidently these carriers are frozen in place when the light is turned off. With each change in \Vbias, the amount of charge needed to screen \Vbias changes in direct proportion. For this reason, and as derived in Appendix B, the shift in \Vturnon is very close to \Vbias, and hence the slope in \figRef{2}{d} is nearly unity. Depending on the required sign of charge to screen the electric field arising from \Vbias and the work function difference, either excess electrons (right inset in \figRef{2}{d}) or excess holes (left inset in \figRef{2}{d}) can be trapped at the interface after illumination.

\subsection{C. Large positive gate biased illumination}

\begin{figure}
\includegraphics[width = \columnwidth]{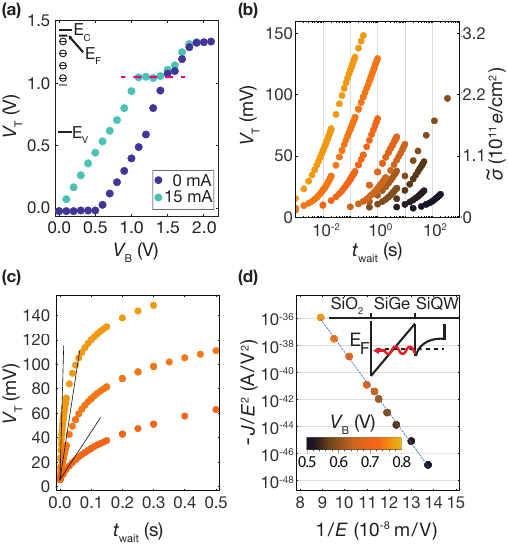}
\caption{
(a) Measured threshold voltage \Vturnon as a function of bias voltage \Vbias after a gate-biased \SI{15}{mA}, \SI{30}{s} illumination (cyan) and a 30s wait with applied bias but with no illumination (purple). The dashed pink line indicates a plateau in the biased illumination data. (b) Measured device turn-on (left axis) and calculated accumulated charge $\sigma$ (right axis) as a function of the duration $t_{\mathrm{wait}}$ under the 9 different values for the applied \Vbias. The colors of the data points match the data shown in (d). (c) Three example data sets from (b) plotted on a linear horizontal axis and shown with a linear fit (black lines) to the short-time data points. These fits are used to acquire the Fowler-Nordheim tunneling current density $J$, which is reported in (d) for all the curves from (b). The inset to (d) shows a schematic band diagram of the Fowler-Nordheim process for this device geometry.
} 
\label{fig3}
\end{figure}

We now explore the effects of biased illumination for larger \Vbias. As shown in \figRef{3}{a}, at \Vbias$=\SI{1.1 \pm 0.1}{V}$, corresponding to \Vturnon$=\SI{1.06 \pm 0.03}{V}$, we observe a plateau  in the threshold voltage after illumination, as indicated by the pink dashed line. This suggests an upper limit to the amount of charge that can be trapped at the oxide interface: as charge fills the available interface states, the Fermi energy increases, eventually crossing into the SiGe conduction band, where any additional added charge at the interface is mobile and can escape through the sample ohmic contacts when the light is turned off.

This saturation suggests an average density of interface states $\bar{D}_{\mathrm{it}} = 2\sigma_{\mathrm{max}}/(eE_G)$, where $E_G = \SI{1.04}{eV}$ is the band gap of Si$_{0.7}$Ge$_{0.3}$~\cite{Schaffler1997}, $e$ is the electron charge, and $\sigma_{\mathrm{max}}$ is the maximum density of trapped charge at the interface, which can be negative above mid-gap or positive below mid-gap. We can estimate $\sigma_{\mathrm{max}}$ as follows: the charge density at the oxide interface required to cancel the electric field in the Si/SiGe during illumination is $\sigma = -\frac{\epsilon_1}{d_1}(V_{\mathrm{B}} - V_{\phi})$ where $\epsilon_1$ and $d_1$ are the dielectric constant and thickness of the gate oxide respectively, 
and $|e|V_{\phi} = \Phi_{\mathrm{Al}} - \chi_{\mathrm{Si}} = \SI{0.23}{eV}$ is the difference between the work function of polycrystalline aluminum, $\Phi_{\mathrm{Al}} = \SI{4.28}{eV}$~\cite{Eastment1973}, and the electron affinity of the silicon quantum well, $\chi_{\mathrm{Si}} = \SI{4.05}{eV}$~\cite{Sze2006}.
We find $\sigma_{\mathrm{max}} = -\frac{\epsilon_1}{d_1}(1.1-0.23)\mathrm{V} = \SI{1.9e12}{e/cm^2}$, and the corresponding average density of interface states $\bar{D}_{\mathrm{it}} = 3.6 \times 10^{12} \mathrm{eV^{-1}cm^{-2}}$.  This value is consistent with the expected magnitude of $\bar{D}_{\mathrm{it}}$ at \SiOx-SiGe interfaces fabricated using low-temperature thermal growth~\cite{Ahn1999}. 

Surprisingly, at even larger \Vbias, we observe a transition in \Vturnon up and out of this plateau. To understand the origin of this transition, we perform an analogous experiment without illumination during the biasing period: we apply a bias voltage \Vbias for 30 seconds \emph{without} any illumination. As expected and as shown in \figRef{3}{a}, for small \Vbias there is no change in \Vturnon. (If there were, the device could not function as an H-FET.) At just above \Vbias$= \SI{0.5}{V}$, \Vturnon begins to shift upward as a function of increasing \Vbias, and, further, the measured \Vturnon joins up smoothly with the measurements of \Vturnon \emph{after} biased illumination. This result strongly suggests that the threshold voltage shifts observed after the plateau shown by the pink dashed horizontal line in \figRef{3}{a} are unrelated to illumination. Evidently, at large enough $V_{\mathrm{B}}$, a charge density can be trapped in excess of $\sigma_{\mathrm{max}}$. 

We now present data showing how this charge accumulates. As shown in \figRef{3}{b}, we find that the measured \Vturnon depends on the amount of time $t_{\mathrm{wait}}$ spent at \Vbias. \red{As with the every measurement of \Vturnon with the H-FET, a reset (biased illumination with \Vbias = 0) is performed in between the data points of Fig.~\ref{fig3} to erase all history of charge accumulation from the previous measurement.} This time  dependence suggests that, during application of large enough \Vbias, a small current flows that enables charge trapping.  We attribute this current to Fowler-Nordheim tunneling, where electrons in the accumulated quantum well tunnel across the SiGe barrier, which becomes triangular in the presence of an applied electric field~\cite{fowler1931,Lu2011,Laroche2015}.  

\red{Interestingly, this process enables shifts in \Vturnon above the observed plateau marked by the pink dashed line in \figRef{3}{a}. We argued above that the plateau arises because the interface states below the band edge have been filled. Electrons that during illumination accumulate at the oxide-semiconductor interface evidently cannot access any additional states during the 30 second illumination.  In contrast, electrons accelerated during the Fowler-Nordheim process have non-zero kinetic energy, and we hypothesize that this energy enables injection of charge into the near-interface region of the oxide or into localized states in the thin silicon cap layer~\cite{Lu2011}, either of which would explain how additional charge is trapped beyond that needed to fill the interface trap states up to the band edge.}

We now extract the Fowler-Nordheim tunneling current density during the initial stages of the experiment. \red{This current density is given by $J = d\tilde{\sigma}(t)/dt$,  where $\tilde{\sigma}(t) = -\frac{\epsilon_1}{d_1}V_{\mathrm{T}}(t)$ is the change in the density of trapped charge arising from $J$, and $\tilde{\sigma}$ is plotted on the right-hand axis of \figRef{3}{b}.} It is important to fit only the first part of the data \Vturnon as a function of $t_{\mathrm{wait}}$, because the accumulation of charge will screen the electric field in the semiconductor, reducing the Fowler-Nordheim current. \figReflong{3}{c} shows such a linear fit to three example curves from \figRef{3}{b}. \red{The Fowler-Nordheim current density is given by,
\begin{equation}
    J(E) = AE^2e^{-B/E} 
    \label{eq1}
\end{equation} 
where the prefactors $A$ and $B$ depend on sample details~\cite{Lenzlinger1969,Ravindra1998}. We extract the familiar $\ln \left( J/E^2 \right)$ vs.~$1/E$ relationship over 11 orders of magnitude, as shown in \figRef{3}{d}. The extracted fit parameters $A=\SI{1.28e-7}{A/V^2}$ and $B=\SI{2.26e8}{V/m}$ predict a minimum bias voltage of $\SI{0.52}{V}$ to generate sufficient tunneling current to shift the threshold voltage beyond our measurement precision (\SI{2}{mV}), in agreement with the onset of Fowler-Nordheim tunneling in \figRef{3}{a}.}

\subsection{D. Large negative gate biased illumination}

\begin{figure}
\includegraphics[width = \columnwidth]{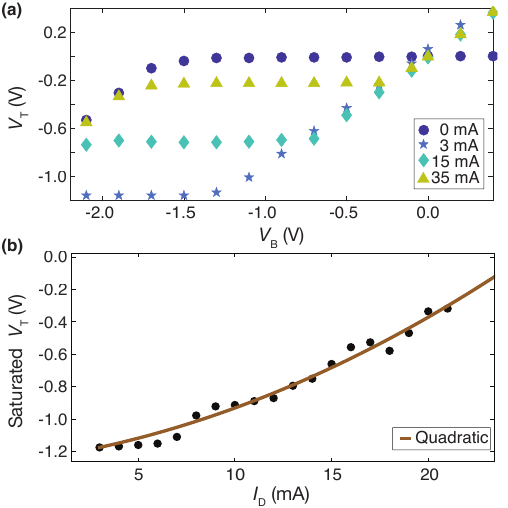}
\caption{
(a) Measured device threshold \Vturnon vs.~illumination bias \Vbias for different laser diode pulse heights in the negative-bias regime. The saturation point becomes power dependent for biases \Vbias$< $ \SI{-0.2}{V}. (b) \red{Measured device threshold vs.~diode pulse height for a fixed pulse time of \SI{30}{s}}. The relationship is found to be quadratic (brown line) with the minimum of the fit at the lasing threshold of \SI{3}{mA}, suggesting a two-photon process.
} 
\label{fig4}
\end{figure}

We now investigate biased illumination for large negative gate voltages. \figReflong{4}{a} shows \Vturnon after biased illumination, for three different laser currents \red{all with a \SI{30}{s} long pulse}, and for the case of no illumination. Unlike experiments done at positive \Vbias, the threshold voltage for negative bias depends sensitively on laser power. Furthermore, we observe a saturation point where \Vbias no longer shifts \Vturnon, and this saturation point is sensitive to the laser power applied during biased illumination.

\figReflong{4}{b} reports the saturation value of the the threshold voltage as a function of the current through the laser diode, which is found to be quadratic with a minimum near the threshold current of the laser diode. The  fact that the saturation value depends on laser intensity suggests that there is a competing process that empties trapped charge at the same time that the screening charge is filling trap states from the electron-hole pairs in the SiGe bands, and the quadratic dependence indicates that such a process may involve two photons. At even larger negative voltages (e.g. \Vbias~$<$~\SI{-1.5}{V} for $I_{\mathrm{D}} = \SI{35}{mA}$), Fowler-Nordheim dynamics again take place, as with the positive bias experiments reported above, and as made evident by comparison with the case of no illumination (purple circles). 

\section{III. Conclusion}
We have shown how the threshold voltage of a device at cryogenic temperatures can be tuned \emph{in-situ} using illumination under an applied gate bias. \red{For low bias, these results are consistent with the simple and intuitive hypothesis that illumination generates electron-hole pairs in the bulk semiconductor, enabling charges to fill states at the interface and screen the electric field arising from the bias voltage applied to the gate.}  These charges are trapped in place after the illumination ends. For larger \Vbias, we have discussed the important roles of the finite density of available interface states, Fowler-Nordheim tunneling, and two-photon charge liberation. These results and the models presented help explain the effectiveness of the widely-used practice of illumination of Si/SiGe quantum dot qubit devices, and they offer possibilities for expanding the use of such illumination to non-zero gate biases, which would not need to be the same on each gate in a device.

\begin{acknowledgments}
We thank HRL Laboratories, LLC for support and L.F. Edge for providing one of the Si/SiGe heterostructure used in this work. We thank Piotr Marciniec for experimental assistance. Research was sponsored in part by the Army Research Office (ARO) under Grant Nos.~W911NF-22-1-0257,  W911NF-17-1-0274, and W911NF-23-1-0110. Development and maintenance of the growth facilities used for fabricating samples were supported by DOE (DE-FG02-03ER46028). We acknowledge the use of facilities supported by NSF through the UW-Madison MRSEC (DMR-1720415) and the NSF MRI program (DMR-1625348). The views and conclusions contained in this document are those of the authors and should not be interpreted as representing the official policies, either expressed or implied, of the Army Research Office (ARO), or the U.S. Government. The U.S. Government is authorized to reproduce and distribute eprints for Government purposes notwithstanding any copyright notation herein.
\end{acknowledgments}

\section{Appendix A: Device Fabrication}
The quantum-dot device shown in \figRef{1}{a} is fabricated on a CVD-grown Si/SiGe heterostructure with a \SI{170}{nm} $\mathrm{Si}_{0.7}\mathrm{Ge}_{0.3}$ relaxed buffer layer, \SI{9}{nm} Si quantum well, \SI{30}{nm} $\mathrm{Si}_{0.7}\mathrm{Ge}_{0.3}$ spacer, and a \SI{1}{nm} Si cap. A \SI{20}{nm} Al$_2$O$_3$ field oxide grown by atomic layer deposition (ALD) isolates the reservoir gates from the implant regions. The gate oxide is grown in the active region of the device where the quantum dots form. This layer consists of \SI{5}{nm} Al$_2$O$_3$, also grown by ALD, to isolate the gate electrodes from the semiconductor. Device bond pads are a \SI{20}{}/\SI{160}{nm} Ti/Pd stack patterned using photo-lithography. The quantum dot gates consists of three overlapping aluminum gate layers with \SI{35}{}/\SI{55}{}/\SI{70}{nm} for the screening, accumulation, and barrier gates respectively, patterned using electron-beam lithography. A $\sim \SI{4}{nm}$ AlO$_x$ intergate oxide is achieved by a plasma-ash oxide enhancement. 

The H-FET measured in this work is fabricated on CVD-grown Si/SiGe heterostructure with a \SI{690}{nm} $\mathrm{Si}_{0.7}\mathrm{Ge}_{0.3}$ buffer layer, \SI{12.5}{nm} Si quantum well, \SI{38}{nm} $\mathrm{Si}_{0.7}\mathrm{Ge}_{0.3}$ spacer, and a \SI{3.4}{nm} Si cap. A gate oxide of thickness approximately $\SI{10}{nm}$ was grown using a \SI{700}{\degree C} dry-oxidation of the Si cap. The \SI{180}{nm} aluminum top gate is patterned using electron-beam lithography. 

\section{Appendix B: Poisson Equation for Si/SiGe with trapped charge}

\begin{figure}
\includegraphics[width = .7\columnwidth]{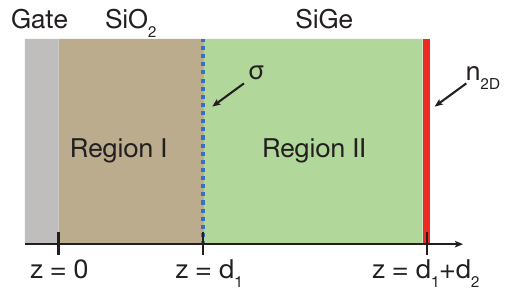}
\caption{
One-dimensional model for the classical electrostatics inside the Si/SiGe devices used in this work.
} 
\label{fig5}
\end{figure}

We show here that a simple one-dimensional model for the Si/SiGe device predicts a shift in \Vturnon equal to the \Vbias applied during illumination. As shown in Fig.~\ref{fig5}, we calculate the classical electrostatics in two regions: the gate oxide with dielectric constant $\epsilon_1$ and thickness $d_1$ and the SiGe spacer with dielectric constant $\epsilon_2$ and thickness $d_2$. Assuming there is no background doping, and making the simplifying assumption that there is no oxide fixed charge, the electrostatic potential, $\phi_i(z)$, in each region $i$ is linear, and the boundary conditions are
\begin{subequations}
\begin{align}
& \phi_{\mathrm{I}}(z=0) = V_{\mathrm{G}} - V_{\phi}\\
& \phi_{\mathrm{I}}(z = d_1) = \phi_{\mathrm{II}}(z = d_1)\\
& \epsilon_1\frac{\partial\phi_{\mathrm{I}}(z)}{\partial z}\Big\rvert_{z = d_1} - \epsilon_2\frac{\partial\phi_{\mathrm{II}}(z)}{\partial z}\Big\rvert_{z = d_1} =  \sigma  \\
& \phi_{\mathrm{II}}(z = d_1+d_2) = 0\\
& \epsilon_2\frac{\partial\phi_{\mathrm{II}}(z)}{\partial z}\Big\rvert_{z = d_1+d_2} = n_{\mathrm{2D}}
\end{align}
\end{subequations}
where $V_{\mathrm{G}}$ is the applied gate voltage and $n_\mathrm{2D}$ is the density of electrons in the quantum well. Eqs.~(2d)-(2e) are consistent with a parallel plate capacitor model where there is no electric field beyond $z > d_1 + d_2$. 

This linear system of equations is solved for $\phi_{\mathrm{I}}(z)$, $\phi_{\mathrm{II}}(z)$, and $V_{\mathrm{G}}$, giving
\begin{subequations}
    \begin{align}
    \phi_{\mathrm{I}}(z) &= V_{\mathrm{G}} - V_{\phi} - \frac{\epsilon_2(V_{\mathrm{G}} - V_{\phi}) - d_2\sigma}{\epsilon_2 d_1 + \epsilon_1 d_2}z\\
    \phi_{\mathrm{II}}(z) &= \frac{\epsilon_1 (V_{\mathrm{G}} - V_{\phi}) - d_1 \sigma}{\epsilon_2 d_1 + \epsilon_1 d_2}\big(z -(d_1+d_2)\big)\\
    V_{\mathrm{G}} &= V_{\phi} - \frac{\epsilon_1 d_2 + \epsilon_2 d_1}{\epsilon_1 \epsilon_2}n_{\mathrm{2D}} - \frac{d_1}{\epsilon_1}\sigma.
    %& n_{\mathrm{2D}} = - \frac{\epsilon_1 \epsilon_2}{\epsilon_1 d_2 + \epsilon_2 d_1}(V_{\mathrm{G}} - V_{\phi}) - \frac{\epsilon_2 d_1 \sigma}{\epsilon_1 d_2 + \epsilon_2 d_1}\\
    \end{align}
\end{subequations}
$V_{\mathrm{T}}$ is the limit of $V_{\mathrm{G}}$ as $n_{\mathrm{2D}}\xrightarrow{}0$. As described in the main text, the charge density at the interface is $\sigma = -\frac{\epsilon_1}{d_1}(V_{\mathrm{B}} - V_{\phi})$. By substituting this charge density into Eq.~(3c) and taking the limit $n_{\mathrm{2D}}\xrightarrow{}0$, it can be seen that $V_{\mathrm{T}} = V_{\mathrm{B}}$ in this model.

\red{As reported above, the slope of the dashed black line in Fig.~2(d) is 0.94, close but not equal to the unity value predicted by this model.  A slope less than unity is consistent with some charge escaping after illumination.  It is possible that during illumination some of the charge that screens the electric field from the gate resides in the conduction band rather than in localized states.  Such charge presumably can escape when the illumination is terminated, consistent with the physical interpretation described above, in Sec.~IIC, of the plateau marked by the pink line in Fig.~3(a).}

\section{Appendix C: Stability and Repeatability}

\begin{figure}
\includegraphics[width = \columnwidth]{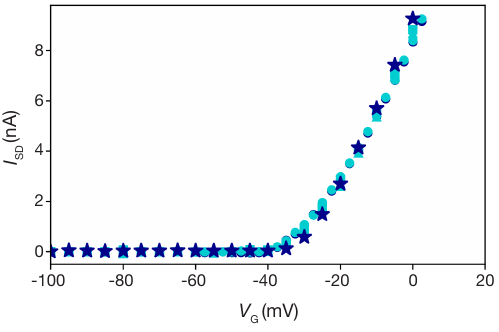}
\caption{
Turn on measurements of an H-FET device following $\SI{0}{V}$ biased illuminations. The teal squares, circles, and triangles report three example turn-on measurements---analogous to those shown in Fig.~2(c)---performed immediately after an illumination with a bias voltage of $\SI{0}{V}$. Blue stars report a measurement performed one month after a $\SI{0}{V}$ biased illumination. 
} 
\label{fig6}
\end{figure}

To demonstrate the level of stability and repeatability of device behavior following illumination, we show in Fig.~\ref{fig6} three turn-on curves, each acquired immediately after $\SI{0}{V}$ biased illuminations (teal squares, circles, and triangle data points). The blue star data points report a turn-on measurement acquired one month after a $\SI{0}{V}$ biased illumination.  Together, these data sets demonstrate the level of stability and repeatability of the turn-on curves following illumination.

%\bibliography{ref}

\providecommand{\noopsort}[1]{}\providecommand{\singleletter}[1]{#1}%

\end{document}